\documentclass[10pt]{article}
\pdfoutput=1
\usepackage{geometry}
\usepackage{amsmath}
\usepackage{mathptmx} %%
\usepackage{graphicx}
\usepackage{hyperref}
\usepackage{cancel}
\usepackage{textcomp}
\usepackage{times}
\usepackage{epsfig}
\usepackage{xcolor}
\usepackage{colortbl}
\usepackage{slashed}
\usepackage{booktabs}
\usepackage{multirow}
\usepackage{latexsym}
\definecolor{mygray}{gray}{.95}
\usepackage[title]{appendix}
\allowdisplaybreaks[4]

\newcommand{\Tr}{\rm Tr}

\newcommand{\hc}{\rm h.c. }
\newcommand{\calL}{{\cal L}}
\newcommand{\calO}{{\cal O}}

\newcommand{\calX}{{\cal X}}
\newcommand{\calY}{{\cal Y}}

\newcommand{\TeV}{\rm TeV}
\newcommand{\GeV}{\rm GeV}
\newcommand{\MeV}{\rm MeV}

\newcommand{\eV}{\rm eV}

\newcommand{\chpt}{\chi{\rm PT}}

\begin{document}
\baselineskip=16pt

\pagenumbering{arabic}

\vspace{1.0cm}

\begin{center}
{\Large\sf Effective field theory approach to lepton number violating decays $K^\pm\rightarrow \pi^\mp l^{\pm}_\alpha l^{\pm}_\beta$: long-distance contribution}
\\[10pt]

\vspace{.5 cm}

{Yi Liao~$^{a,c}$\footnote{liaoy@nankai.edu.cn},~
Xiao-Dong Ma~$^{b}$\footnote{maxid@phys.ntu.edu.tw},~
Hao-Lin Wang~$^{a}$\footnote{whaolin@mail.nankai.edu.cn} }

{$^a$~School of Physics, Nankai University, Tianjin 300071, China
\\
$^b$ Department of Physics, National Taiwan University, Taipei 10617, Taiwan
\\
$^c$ Center for High Energy Physics, Peking University, Beijing 100871, China}

\vspace{2.0ex}
{\bf Abstract}
\end{center}

This is a sequel to our recent work~\cite{Liao:2019gex} in which we calculated the lepton number violating (LNV) $K^\pm$ decays due to contact dimension-9 (dim-9) quark-lepton effective interactions that are induced at a high energy scale. In this work we investigate the long-distance contribution to the decays arising from the exchange of a neutrino. These decays can probe LNV interactions involving the second generation of fermions that are not reachable in nuclear neutrinoless double-$\beta$ decays. Our study is completely formulated in the framework of effective field theories (EFTs), from the standard model effective field theory (SMEFT) through the low energy effective field theory (LEFT) to chiral perturbation theory ($\chpt$). We work to the first nontrivial orders in each effective field theory, collect along the way the matching conditions and renormalization group effects, and express the decay branching ratios in terms of the Wilson coefficients associated with the dim-5 and dim-7 operators in SMEFT. Our result is general in that it does not depend on dynamical details of physics at a high scale that induce the effective interactions in SMEFT and in that it does not appeal to any hadronic models. We find that the long-distance contribution overwhelmingly dominates over the contact or short-distance one. Assuming the new physics scale to be around a TeV, the branching ratios are predicted to be below the current experimental upper bounds by several orders of magnitude.

\newpage
%%%%%%%%%%%%%%%%%%%%%%%
\section{Introduction}
%%%%%%%%%%%%%%%%%%%%%%%

The origin of neutrino mass and the nature of neutrinos remain a challenging issue in physics beyond the standard model. If neutrinos are Majorana fermions, the lepton number is violated by two units. In that case it is desirable to explore lepton number violating (LNV) signals beyond the Majorana neutrino masses. At a high energy collider such as the LHC, the LNV signals usually manifest themselves as like-sign multileptons that supposedly originate from the decays of new heavy particles engaged in neutrino mass generation~\cite{Khachatryan:2015gha}. The null search result then sets a lower bound on the masses of new particles under some simplifying assumptions. Complementary to these direct searches are high-precision experiments at low energy that seek the imprints of new physics in rare or forbidden processes. The most extensively studied so far are the so-called nuclear neutrinoless double $\beta$ ($0\nu\beta\beta$) decays, $X\to X'e^\mp e^\mp$, in which a parent nucleus $X$ decays into a daughter nucleus $X'$ with the release of a pair of like-sign electrons or positrons~\cite{Furry:1939qr,Barabash:2011mf}. The null result in current experiments can be then used to set a strong bound on the relevant LNV physics~\cite{KamLAND-Zen:2016pfg,GERDA:2018zzh}.

Other nuclear processes proposed to search for include, for instance, the muon to positron or antimuon conversion $\mu^-X\rightarrow e^+(\mu^+)X^\prime$ in the upcoming Mu2e experiment~\cite{Bernstein:2019fyh}. On the other hand, there is a plethora of flavor physics experiments in recent years that search for LNV decays in flavored and charged mesons such as $K^\pm,~D^\pm,~D^\pm_s,~B^\pm$ and the $\tau$ lepton~\cite{Amhis:2016xyh,
CortinaGil:2019dnd,Appel:2000tc,Aaij:2013sua,Rubin:2010cq,Lees:2011hb,
Kodama:1995ia,Aaij:2014aba,BABAR:2012aa,Lees:2013gdj,Aaij:2011ex,
Aaij:2012zr,Seon:2011ni,Miyazaki:2012mx}, and the bounds on some of the decays are expected to be considerably improved in future experiments~\cite{Chun:2019nwi,Perez:2019cdy}.

From the theoretical point of view the LNV decays of the flavored mesons and the $\tau$ lepton are sensitive to the effective interactions of the fermions beyond the first generation that cannot be probed in the nuclear $0\nu\beta\beta$ decays due to kinematical limitations, and can thus provide complementary information on underlying new physics. These decays can be best investigated with the aid of various effective theories while avoiding theoretical uncertainties associated with nuclear physics. In a recent publication~\cite{Liao:2019gex} we started the endeavor with the decays $K^\pm\to\pi^\mp l^\pm l^\pm$ arising from effective contact interactions among light quarks and charged leptons $l=e,~\mu$. In this work we make a comprehensive analysis on the decays by incorporating the long-distance contribution due to the exchange of neutrinos. Before diving into technical details we describe briefly the strategy of our study in the framework of effective field theory (EFT).

\begin{figure}[!h]
\centering
\includegraphics[width=15cm]{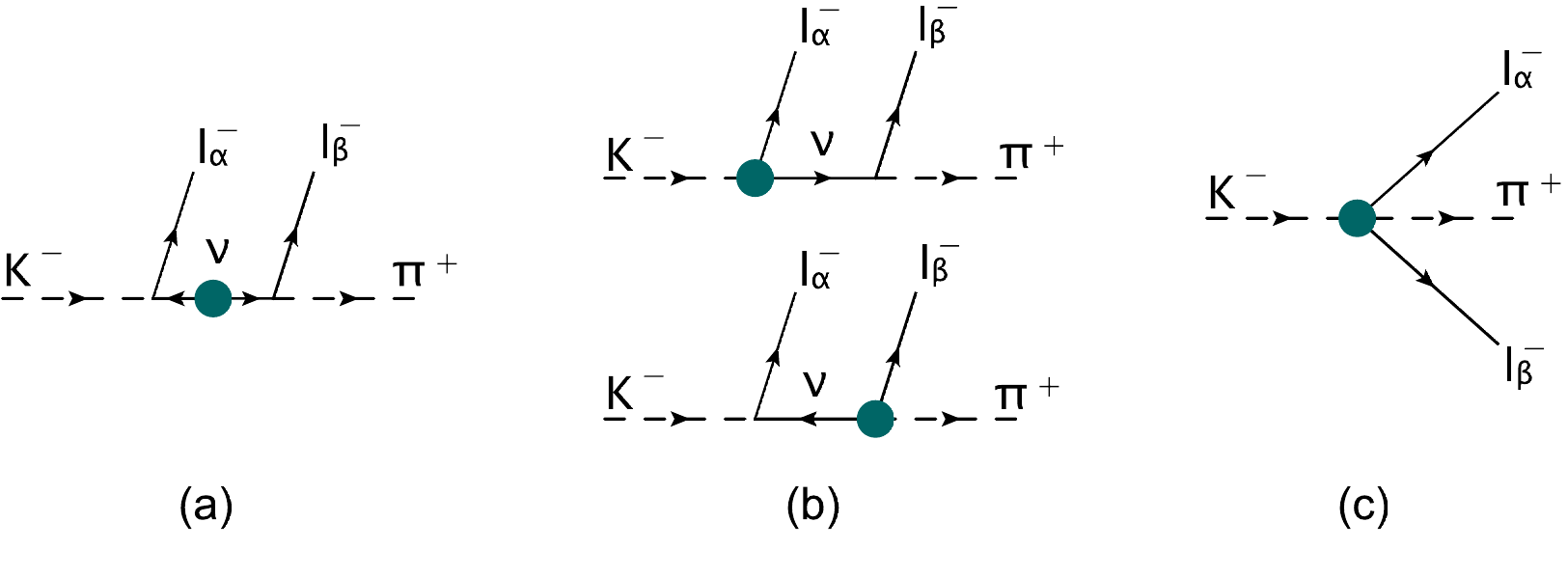}
\caption{Generic Feynman diagrams for the decay $K^-\rightarrow \pi^+l^-_\alpha l^-_\beta$ in $\chpt$, where the heavy blob stands for effective LNV interactions. Diagrams with the two charged leptons crossed are not shown in (a, b).}
\label{fig1}
\end{figure}

In the low energy region defined by the kaon and pion masses, the relevant dynamical degrees of freedom are the octet of the pseudo-Nambu-Goldstone bosons ($\pi,~K,~\eta$), charged leptons, neutrinos, and the photon, if we assume there are no new very light particles. The low-energy manifestations of lepton number violation from any high-scale new physics are reflected in the effective interactions of those light particles, which can be systematically organized in chiral perturbation theory ($\chpt$) formulated in terms of external sources~\cite{Gasser:1983yg,Gasser:1984gg,Cata:2007ns}. In particular, working to the leading order in $\chpt$ and in LNV effects, lepton number violation could manifest itself through the neutrino mass (shown in figure~\ref{fig1}(a)), effective interactions of a single meson with a charged lepton and neutrino pair (figure~\ref{fig1}(b)) or of the two mesons with a pair of likely-charged leptons (figure~\ref{fig1}(c)). The short-distance (SD) contribution in figure~\ref{fig1}(c) was thoroughly studied in~\cite{Liao:2019gex}, and this work will focus on the long-distance (LD) terms in figure~\ref{fig1}(a, b) due to the exchange of neutrinos. It is easy to parameterize the above effective interactions, but our aim is to work them out systematically in the EFT approach by matching sequentially to EFTs closer to new physics at a high scale. In this manner we are able to express the decay branching ratios in terms of the Wilson coefficients in the EFT defined at the electroweak scale or an even higher scale when necessary.

The paper is organized as follows. We start in section~\ref{sec2} with the low energy effective field theory (LEFT) defined between the electroweak scale $\Lambda_\textrm{EW}$ and chiral symmetry breaking scale $\Lambda_\chi$~\cite{Buchalla:1995vs,Jenkins:2017jig}. We collect the relevant dimension-3 (dim-3), -6 and -7 LNV effective operators and discuss their one-loop QCD running effects. (The dim-3, -4, and -5 operators are the mass, kinetic, and electromagnetic (transition) moment terms.) We then match them to other EFTs along the ladder of scales. In section~\ref{sec3}, the dim-6 and -7 operators in LEFT are matched at the scale $\Lambda_\chi$ to $\chpt$, thus determining the vertices in figure~\ref{fig1} in terms of the Wilson coefficients in LEFT and low energy constants (LECs) of QCD strong dynamics. Then in section~\ref{sec4} we match LEFT upwards the scale to the standard model effective field theory (SMEFT), assuming that there are no new particles with a mass of order $\Lambda_\textrm{EW}$ or lower. The relevant leading LNV operators in SMEFT are the dim-5 and -7 ones known in the literature~\cite{Weinberg:1979sa, Lehman:2014jma,Liao:2016hru,Liao:2019tep} and reproduced in appendix~\ref{app1}. We express the branching ratios in terms of the Wilson coefficients in SMEFT and the LECs of strong dynamics through the above two-step matching, whose complete expressions can be found in appendix~\ref{app2}, and make some numerical estimates. Our main findings are finally recapitulated in section~\ref{sec5}.

%%%%%%%%%%%%%%%%%%%%%%%
\section{LNV effective interactions in LEFT}
%%%%%%%%%%%%%%%%%%%%%%%
\label{sec2}

The low energy effective field theory is an EFT defined between the electroweak scale $\Lambda_\textrm{EW}\sim 10^2~\GeV$ and chiral symmetry breaking scale $\Lambda_\chi=4\pi F_\pi\sim 1~\GeV$.
The dynamical degrees of freedom include five quarks ($u,~d,~s,~c,~b$), all charged leptons ($e,~\mu,~\tau$) and neutrinos ($\nu_e,~\nu_\mu,~\nu_\tau$), the photon and gluons, and enjoy QED and QCD gauge symmetries. LEFT has been fruitfully applied particularly in flavor physics~\cite{Buchalla:1995vs}. As a low energy effective theory, it is an organized and infinite series of effective interactions whose importance is relatively measured by canonical dimensions of effective operators with similar symmetry properties.
For our purpose here we focus on the effective operators that violate the lepton number $L$ by two units and potentially contribute to the decays under consideration at the leading orders.

We work without losing generality in the convention that quarks and charged leptons have been diagonalized while neutrinos are in their flavor states. Different conventions amount to different ways to incorporate quark and lepton mixing matrices in generally unknown Wilson coefficients. We first recall that the SD contribution in figure~\ref{fig1}(c) arises at leading order from dim-9 operators involving two like-sign leptons and four quarks which have been thoroughly analyzed in Ref.~\cite{Liao:2019gex}. In the following we investigate systematically the effective operators that could finally dominate the LD contribution in figure~\ref{fig1}(a,b). The dim-3 operator is unique, i.e., the Majorana neutrino mass term in the effective Lagrangian:
\begin{eqnarray}
\label{nu_mass}
-\frac{1}{2}m_{\alpha\beta}\overline{\nu^C_\alpha}\nu_\beta,
\label{numass}
\end{eqnarray}
where $m_{\alpha\beta}$ is the neutrino mass matrix in the flavor basis of the neutrinos $\nu_\alpha=\nu_e,~\nu_\mu,~\nu_\tau$ and the superscript $C$ refers to charge conjugation. For the LNV interactions in figure~\ref{fig1}(b), the relevant operators in LEFT involve one charged lepton, one neutrino, and a pair of quarks. These operators first appear at dimension six and have been classified in Ref.~\cite{Jenkins:2017jig}. Following our notations in~\cite{Liao:2019gex}, we denote them as follows:
\begin{align}
\label{ORLS}
\calO^{RL,S}_{pr\alpha\beta}&=
(\overline{u_R^p}d_L^r)(\overline{l_{L\alpha}}\nu^C_\beta), &
\calO^{LR,S}_{pr\alpha\beta}&=
(\overline{u_L^p}d_R^r)(\overline{l_{L\alpha}}\nu^C_\beta),
\\
\calO^{LL,V}_{pr\alpha\beta}&=
(\overline{u_L^p}\gamma_\mu d_L^r)(\overline{l_{R\alpha}}\gamma^\mu \nu^C_\beta), &
\calO^{RR,V}_{pr\alpha\beta}&=
(\overline{u_R^p}\gamma_\mu d_R^r)(\overline{l_{R\alpha}}\gamma^\mu \nu^C_\beta),
\\
\calO^{LR,T}_{pr\alpha\beta}&=
(\overline{u_L^p}\sigma_{\mu\nu}d_R^r)
(\overline{l_{L\alpha}}\sigma^{\mu\nu}\nu^C_\beta).
\end{align}
Here the Latin letters $p,r$ indicate the flavors of the up- and down-type quarks $u^p,~d^r$ whose chiralities ($L,~R$) are shown by their subscripts and by the first two superscripts of the operators $\calO$. Since $\nu_\beta^C$ is right-handed, the chirality of the charged lepton field in a lepton bilinear is automatically determined by the type of the operators, $S,~V,~T$. In addition we also require the SM effective operators due to charged-current interactions between leptons and quarks:
\begin{eqnarray}
\calO^\textrm{(SM)}_{pr\alpha\beta}
=V_{pr}(\overline{u_L^p}\gamma_\mu d_L^r)
(\overline{l_{L\alpha}}\gamma^\mu\nu_\beta)\delta^{\alpha\beta},
\end{eqnarray}
where the $V_{pr}$ is the Cabibbo-Kobayashi-Maskawa (CKM) matrix.

At the next order LNV operators carry a covariant derivative $D_\mu$ to become dimension 7. Considering the restrictions and reductions due to gauge symmetry, equations of motion, integration by parts, and Fierz identities, we obtain the following LNV operators relevant to our purpose here~\cite{LiaoMaWang2020}:
\begin{align}
\calO^{LL,VD}_{pr\alpha\beta}&=(\overline{u_L^p}\gamma_\mu d_L^r)(\overline{l_{L\alpha}}i\overleftrightarrow{D}^\mu\nu^C_\beta), &
\calO^{RR,VD}_{pr\alpha\beta}&=(\overline{u_R^p}\gamma_\mu d_R^r)(\overline{l_{L\alpha}}i\overleftrightarrow{D}^\mu\nu^C_\beta),
\\
\calO^{LR,TD}_{pr\alpha\beta}&=
(\overline{u_L^p}\sigma_{\mu\nu}d_R^r)(\overline{l_{R\alpha}}\gamma^{[\mu} \overleftrightarrow{D}^{\nu]}\nu^C_\beta), &
\calO^{RL,TD}_{pr\alpha\beta}&=
(\overline{u_R^p}\sigma_{\mu\nu}d_L^r)(\overline{l_{R\alpha}}\gamma^{[\mu} \overleftrightarrow{D}^{\nu]}\nu^C_\beta),
\label{ORLTD}
\end{align}
where $\overline{A}\overleftrightarrow{D}^\mu B=\overline{A}(D^\mu B)-\overline{A}\overleftarrow{D^\mu}B$ and $\gamma^{[\mu}D^{\nu]}=\gamma^\mu D^\nu-\gamma^\nu D^\mu$. The operators in equations~\eqref{ORLS}-\eqref{ORLTD}, as well as the dim-3 Majorana mass term \eqref{numass}, make up the main body for the LD contribution. These operators will be matched in the next section to those in $\chpt$ where the lepton bilinears act as external sources.

Since we will match the effective interactions in LEFT to those in SMEFT at the scale $\Lambda_{\text{EW}}$ and to those in $\chpt$ at the scale $\Lambda_\chi$, it is necessary to sum the large logarithms between the two scales using renormalization group equations. In our case the leading effect arises from the 1-loop QCD renormalization. While the vector-type operators are free of renormalization, the scalar- and tensor-type operators are indeed renormalized, whose Wilson coefficients satisfy the renormalization group equations:
\begin{eqnarray}
&&\mu{d \over d\mu}C^S=-{\alpha_s \over 2\pi}3C_FC^S,~~~
C^S\in \Big\{C_{pr\alpha\beta}^{RL,S},C_{pr\alpha\beta}^{LR,S}\Big\},
\\
&&\mu{d \over d\mu}C^T={\alpha_s \over 2\pi}C_FC^T,~~~
C^T\in \Big\{C_{pr\alpha\beta}^{LR,T},~C_{pr\alpha\beta}^{LR,TD},~
C_{pr\alpha\beta}^{RL,TD}\Big\},
\end{eqnarray}
where $C_F=(N^2-1)/(2N)=4/3$ with $N=3$ being the color number. The solutions between the scales $\mu_1$ and $\mu_2$ are
\begin{align}
C^S(\mu_1)&=\left({ \alpha_s(\mu_2)\over \alpha_s(\mu_1)} \right)^{3C_F/b}C^S(\mu_2), &
C^T(\mu_1)&=\left({ \alpha_s(\mu_2)\over \alpha_s(\mu_1)} \right)^{-C_F/b}C^T(\mu_2),
\end{align}
where $b=-11+2n_f/3$ with $n_f$ being the number of active quark flavors. Incorporating quark threshold effects, we obtain the numerical results between the scales $\Lambda_\chi$ and $\Lambda_{\text{EW}}$:
\begin{align}
\label{rge1}
C^S(\Lambda_\chi)&=1.656 C^S(\Lambda_\text{EW}), &
C^T(\Lambda_\chi)&=0.845C^T(\Lambda_\text{EW}).
\end{align}
Thus the scalar-type interactions are enhanced while the tensor-type ones are suppressed when evolving down from the high scale $\Lambda_\text{EW}$ to the low scale $\Lambda_\chi$.

%%%%%%%%%%%%%%%%%%%%%%%
\section{Matching onto effective interactions in $\chpt$}
%%%%%%%%%%%%%%%%%%%%%%%
\label{sec3}

While the charged leptons and neutrinos retain their identities at low energy, the quark and gluon degrees of freedom will condense into hadrons due to strong dynamics. Since the process $K^-\rightarrow\pi^+l^-_\alpha l^-_\beta$ in question involves only the light quarks $q=u,~d,~s$, its transition matrix element due to effective interactions in LEFT can be beautifully evaluated by matching to chiral perturbation theory, which is the low energy effective field theory of QCD. $\chi$PT is based on the fact that the QCD Lagrangian has the approximate chiral symmetry $SU(3)_L\times SU(3)_R$ for the three light quarks which is spontaneously broken by the quark condensate $\langle\bar qq\rangle=-3BF_0^2$ to the diagonal $SU(3)_V$. The symmetry breakdown brings about eight pseudo-Nambu-Goldstone bosons (pNGBs), which are identified with the octet of the lowest-lying pseudoscalars $\pi^\pm,~\pi^0,~K^\pm,~K^0,~\overline{K^0},~\eta$. In the $\chpt$ formalism they are represented by the element in the coset space $SU(3)_L\times SU(3)_R/SU(3)_V$ and take the matrix form,
\begin{align}
U(x)&=\exp\left(\frac{i\sqrt{2}\Pi(x)}{F_0}\right), &
\Pi&=\begin{pmatrix}
\frac{\pi^0}{\sqrt{2}}+\frac{\eta}{\sqrt{6}} & \pi^+ & K^+
\\
\pi^- & -\frac{\pi^0}{\sqrt{2}}+\frac{\eta}{\sqrt{6}} & K^0
\\
K^- & \bar{K}^0 & -\sqrt{\frac{2}{3}}\eta
\end{pmatrix},
\end{align}
where $F_0$ is the decay constant in the chiral limit. Corresponding to chiral transformations of quarks $q_L\to Lq_L$ and $q_R\to Rq_R$, $U$ transforms as $U\rightarrow LUR^\dagger$ with $L\in SU(3)_L$ and $R\in SU(3)_R$.

The interactions of pNGBs with leptons due to dim-6 and dim-7 operators in equations~\eqref{ORLS}-\eqref{ORLTD} can be realized through the external source method in which the global chiral symmetry is promoted to a local one~\cite{Gasser:1983yg,Gasser:1984gg,Cata:2007ns}.
At the quark-gluon level, the QCD Lagrangian with all possible external sources is parameterized as follows,
\begin{eqnarray}
\label{external}
\mathcal{L}=\mathcal{L}_\textrm{QCD}
+\overline{q_L}l_\mu \gamma^\mu q_L+\overline{q_R}r_\mu \gamma^\mu q_R
+\left[\overline{q_L}(s-ip)q_R
+\overline{q_L}t_l^{\mu\nu}\sigma_{\mu\nu}q_R+\hc\right],
\end{eqnarray}
where $\mathcal{L}_\textrm{QCD}$ is the QCD Lagrangian for massless $u,~d,~s$ quarks. The external sources, $l_\mu=l_\mu^\dagger$, $r_\mu=r_\mu^\dagger$, $s=s^\dagger$, $p=p^\dagger$, $t_r^{\mu\nu}=t_l^{\mu\nu\dagger}$, are $3\times 3$ matrices in flavor space, and transform under chiral group as $l_{\mu}\to L l_{\mu} L^{\dagger}+i L \partial_{\mu} L^{\dagger}$, $r_{\mu}\to R r_{\mu} R^{\dagger}+i R \partial_{\mu} R^{\dagger}$, $\chi\to L\chi R^{\dagger}$, $t_l^{\mu\nu}\to Lt_l^{\mu\nu} R^{\dagger}$, where $\chi=2B(s-ip)$. By comparing the external sources in equation~\eqref{external} with the effective interactions in LEFT formed with the operators in equations~(\ref{ORLS}-\ref{ORLTD}) multiplied by their Wilson coefficients, one singles out the terms in external sources specific to the $K^-\to\pi^+$ transition:
\begin{eqnarray}
(l^\mu)_{ui}&=&-2\sqrt{2}G_FV_{ui}
(\overline{l_{L\alpha}}\gamma^\mu\nu_\alpha)
\nonumber
\\
&&+C^{LL,V}_{ui\alpha\beta}(\overline{l_{R\alpha}}\gamma^\mu\nu^C_\beta)
+C^{LL,VD}_{ui\alpha\beta}(\overline{l_{L\alpha}}
i\overleftrightarrow{D}^\mu\nu^C_\beta)+\cdots,
\\
(r^\mu)_{ui}&=&C^{RR,V}_{ui\alpha\beta}
(\overline{l_{R\alpha}}\gamma^\mu \nu^C_\beta)
+C^{RR,VD}_{ui\alpha\beta}(\overline{l_{L\alpha}}
i\overleftrightarrow{D}^\mu\nu^C_\beta)+\cdots,
\\
(\chi^\dagger)_{ui} &=&
C^{RL,S}_{ui\alpha\beta}(\overline{l_{L\alpha}}\nu^C_\beta)+\cdots,
\\
(\chi)_{ui} &=&
C^{LR,S}_{ui\alpha\beta}(\overline{l_{L\alpha}}\nu^C_\beta)+\cdots,
\\
(t_l^{\mu\nu})_{ui}&=&
C^{LR,T}_{ui\alpha\beta}(\overline{l_{L\alpha}}\sigma^{\mu\nu}\nu^C_\beta)
+C^{LR,TD}_{ui\alpha\beta}(\overline{l_{R\alpha}}\gamma^{[\mu} \overleftrightarrow{D}^{\nu]}\nu^C_\beta)+\cdots,
\\
(t_r^{\mu\nu})_{ui}&=&
C^{RL,TD}_{ui\alpha\beta}(\overline{l_{R\alpha}}\gamma^{[\mu} \overleftrightarrow{D}^{\nu]}\nu^C_\beta)+\cdots,
\end{eqnarray}
where $i$ can be either $d$ or $s$ quark, and the ellipsis denotes terms not relevant to the transition. In $\chpt$ the vector and scalar sources already appear at order $\calO(p^2)$~\cite{Gasser:1983yg,Gasser:1984gg}
\begin{eqnarray}
\label{p2l}
\mathcal{L}^{(2)}_{\chpt}
=\frac{F_0^2}{4}{\Tr}\left(D_\mu U (D^\mu U)^\dagger \right)+\frac{F_0^2}{4}{\Tr} \left(\chi U^\dagger +U\chi^\dagger \right),
\end{eqnarray}
where
\begin{eqnarray}
D_\mu U=\partial_\mu U-i l_\mu U+i U r_\mu,
\end{eqnarray}
while the tensor sources first appear at $\calO(p^4)$~\cite{Cata:2007ns}
\begin{eqnarray}
\label{p4l}
\mathcal{L}^{(4)}_{\chpt}\supset i\Lambda_2 {\Tr}\left(t_l^{\mu\nu}(D_\mu U)^\dagger U (D_\nu U)^\dagger+t_r^{\mu\nu}D_\mu UU^\dagger D_\nu U\right),
\end{eqnarray}
where $\Lambda_2$ is a low energy constant (LEC). Since the tensor structure in equation~\eqref{p4l} involves at least two pNGBs, one charged and one neutral, it cannot contribute at tree level to the process under consideration and will be ignored below.
The expansion of equation~\eqref{p2l} yields the following terms relevant to the LD contribution to the decay $K^-\rightarrow\pi^+l^-_\alpha l^-_\beta$:
\begin{eqnarray}
\calL^{(2)}_{\chpt}&\supset&
F_0\Big[ G_F\left(V_{ud}\partial_\mu \pi^- +V_{us}\partial_\mu K^-\right)
\left(\overline{l_{L\alpha}}\gamma^\mu\nu_\alpha\right)
\nonumber
\\
&&~~~~~
+iB\left( c_{\pi 1}^{\alpha\beta} \pi^-+c_{K1}^{\alpha\beta}K^-\right)
\left(\overline{l_{L\alpha}}\nu^C_\beta\right)
\nonumber
\\
&&~~~~~
-\left( c_{\pi 2}^{\alpha\beta}\partial_\mu \pi^-+c_{K2}^{\alpha\beta}\partial_\mu K^-\right)
\left(\overline{l_{R\alpha}}\gamma^\mu\nu^C_\beta\right)
\nonumber
\\
&&~~~~~
-\left( c_{\pi 3}^{\alpha\beta}\partial_\mu \pi^-+c_{K3}^{\alpha\beta}\partial_\mu K^-\right) \left(\overline{l_{L\alpha}}i\overleftrightarrow{D}^\mu \nu^C_\beta\right)\Big],
\label{llong}
\end{eqnarray}
where the parameters defined at the scale $\Lambda_\chi$ are
\begin{eqnarray}
c_{P_i1}^{\alpha\beta}&=&
\frac{\sqrt{2}}{2}
\left(C^{RL,S}_{ui\alpha\beta}-C^{LR,S}_{ui\alpha\beta}\right), 
\nonumber
\\
c_{P_i2}^{\alpha\beta}&=&
\frac{\sqrt{2}}{4}\left(C^{LL,V}_{ui\alpha\beta}-C^{RR,V}_{ui\alpha\beta} \right),
\nonumber
\\
c_{P_i3}^{\alpha\beta}&=&
\frac{\sqrt{2}}{4}\left(C^{LL,VD}_{ui\alpha\beta}-C^{RR,VD}_{ui\alpha\beta} \right),
\label{cpi1}
\end{eqnarray}
with $P_i=\pi,~K$ for $i=d,~s$ (and sometimes $i=1,~2$). We note in passing that the leading LD contribution does not introduce new LECs of QCD strong dynamics. The above results show that the dim-6 vector-type operators are suppressed by $\calO(p/B)$ relative to their scalar-type counterparts while dim-7 vector-type operators are further suppressed by $\calO(p/\Lambda_{\text{EW}})$. To put it in short, among the dim-6 and -7 LNV operators in LEFT, the scalar-type dim-6 operators generically dominate the LD contribution.

%%%%%%%%%%%%%%%%%%%%%%%
\section{Matching onto SMEFT and decay branching ratios}
%%%%%%%%%%%%%%%%%%%%%%%
\label{sec4}

Now we make connections between the effective interactions in SMEFT and LEFT, so that we can parameterize the decay branching ratios as a function of the SMEFT Wilson coefficients. The dim-5 and -7 LNV operators in SMEFT are reproduced in appendix~\ref{app1} where we slightly improve the basis of dim-7 operators over Ref.~\cite{Liao:2016hru}. At the scale $\Lambda_{\text{EW}}$ where the electroweak symmetry spontaneously breaks down, we integrate out the heavy SM particles $(W,~Z,~h,~t)$ to induce effective interactions in LEFT. The matching results at $\Lambda_{\text{EW}}$ for the Wilson coefficients of the relevant dim-3, -6, and -7 operators in LEFT are, in terms of those of the dim-5 and -7 operators in SMEFT,
\begin{align}
m_{\alpha\beta}&=-v^2C_{LH5}^{\alpha\beta*}
-\frac{1}{2} v^4C_{LH}^{\alpha\beta*}, &
C^{LR,T}_{pr\alpha\beta}&=
\frac{v}{\sqrt{2}}C_{\bar{d}QLLH2}^{rp\alpha\beta*},
\\
C^{RL,S}_{pr\alpha\beta}&=
\frac{v}{\sqrt{2}}V_{wr}C_{\bar{Q}uLLH}^{wp\alpha\beta*}, &
C^{LR,S}_{pr\alpha\beta}&=
\frac{v}{\sqrt{2}}C_{\bar{d}QLLH1}^{rp\alpha\beta*},
\\
C^{LL,V}_{pr\alpha\beta}&=
\frac{v}{\sqrt{2}}V_{pr}C_{LeHD}^{\beta\alpha *}, &
C^{RR,V}_{pr\alpha\beta}&=
\frac{v}{\sqrt{2}}C_{\bar{d}uLeH}^{r p\beta\alpha*}, 
\\
C^{LL,VD}_{pr\alpha\beta}&=
V_{pr}\left( 4C_{LHW}^{\alpha\beta*}+C_{LDH1}^{\alpha\beta*}\right), &
C^{RR,VD}_{pr\alpha\beta}&=2C_{\bar{d}uLDL}^{rp\alpha\beta*},
\end{align}
where $v\approx 246~\GeV$ is the vacuum expectation value of the Higgs field and we have neglected contributions suppressed by small Yukawa couplings. Incorporating the 1-loop QCD running effect in equation~(\ref{rge1}), the $c_i$ parameters in equation~\eqref{llong} defined at $\Lambda_\chi$ are expressed in terms of the SMEFT Wilson coefficients defined at $\Lambda_{\text{EW}}$:
\begin{align}
\label{wcpi1}
c_{P_i1}^{\alpha\beta}&=
\frac{v}{2}(1.656)\calY^{\alpha\beta}_{P_i1}, &
c_{P_i2}^{\alpha\beta}&=\frac{v}{4}\calY^{\alpha\beta}_{P_i2}, &
c_{P_i3}^{\alpha\beta}&=\frac{\sqrt{2}}{4}\calY^{\alpha\beta}_{P_i3},
\end{align}
where
\begin{eqnarray}
\calY^{\alpha\beta}_{P_i1}&=&
V_{wi}C_{\bar{Q}uLLH}^{w1\alpha\beta*}-C_{\bar{d}QLLH1}^{i1\alpha\beta*}, \nonumber
\\
\calY^{\alpha\beta}_{P_i2}&=&
V_{ui}C_{LeHD}^{\beta\alpha*}
-C_{\bar{d}uLeH}^{i1\beta\alpha*}, 
\nonumber
\\
\calY^{\alpha\beta}_{P_i3}&=&
V_{ui}\left(4C_{LHW}^{\alpha\beta*}+C_{LDH1}^{\alpha\beta*}\right)
-2C_{\bar{d}uLDL}^{i1\alpha\beta*}.
\end{eqnarray}

We are now in a position to employ equation~\eqref{llong} to calculate the LD contribution to the $K^-$ decay shown in figure~\ref{fig1}(a,b). To make our answer complete, we include the SD contribution in figure~\ref{fig1}(c) which takes the form~\cite{Liao:2019gex}
\footnote{The convention of $c_i$ differs from that in Ref.~\cite{Liao:2019gex} by a factor $2F_0G_F^2$.},
\begin{eqnarray}
\label{lshort}
\frac{\calL_\textrm{SD}}{F_0^2G_F}
=c_1^{\alpha\beta}K^-\pi^-\overline{l_{L\alpha}}l^C_{L\beta}
+c_5^{\alpha\beta}\partial^\mu K^-\partial_\mu\pi^-\overline{l_{L\alpha}}l^C_{L\beta},
\end{eqnarray}
where
\begin{eqnarray}
c_1^{\alpha\beta}&=&
-2\sqrt{2}\left(0.62 g_{8\times 8}^a+0.88g_{8\times 8}^b\right)
\calX^{\alpha\beta}_1, 
\nonumber
\\
c_5^{\alpha\beta}&=&
-2\sqrt{2}(1.3g_{27\times1})V_{ud}V_{us}
\calX^{\alpha\beta}_2.
\label{wc1}
\end{eqnarray}
The $\calX$ parameters are sums of the Wilson coefficients of the dim-7 operators in SMEFT defined at $\Lambda_{\text{EW}}$,
\begin{eqnarray}
\calX^{\alpha\beta}_1&=&
2\left(V_{us}C_{\bar{d}uLDL}^{11\alpha\beta*}
+V_{ud}C_{\bar{d}uLDL}^{21\alpha\beta*}\right), 
\nonumber
\\
\calX^{\alpha\beta}_2&=&
2C_{LHW}^{\alpha\beta*}+2C_{LHW}^{\beta\alpha*}
+2C_{LDH1}^{\alpha\beta*}+C_{LDH2}^{\alpha\beta*},
\end{eqnarray}
while the QCD LECs determined in~\cite{Cirigliano:2017ymo} are, in our notation~\cite{Liao:2019gex}, $g_{27\times 1}=0.38\pm 0.08$,
$g_{8\times 8}^a=(5.5\pm 2)~{\GeV}^2$, and $g_{8\times 8}^b=(1.55\pm 0.65)~{\GeV}^2$. The complete amplitude for the decay $K^-(k)\rightarrow\pi^+(p)l^-_\alpha(p_1)l^-_\beta(p_2)$ is,
\begin{eqnarray}
\frac{\mathcal{M}}{F_0^2G_F}= T_\textrm{SD}\overline{u_\alpha}P_Ru_\beta^C
+T_{1\mu\nu}\overline{u_\alpha}\gamma^\mu\gamma^\nu P_Ru_\beta^C
+T_{2\mu\nu\rho}\overline{u_\alpha}\gamma^\mu\gamma^\nu\gamma^\rho P_Ru_\beta^C
+T_{3\mu\nu\rho}\overline{u_\alpha}\gamma^\mu\gamma^\nu\gamma^\rho P_Lu_\beta^C,
\label{amplitude}
\end{eqnarray}
where $T_\textrm{SD}$ stands for the SD term and the others are the LD ones:
\begin{eqnarray}
T_\textrm{SD}&=&2c_1^{\alpha\beta}+2c_5^{\alpha\beta} k\cdot p,
\\
\nonumber
T_{1\mu\nu}&=&G_FV_{ud}V_{us}m_{\alpha\beta}\left( k_\mu p_\nu t^{-1}
+p_\mu k_\nu u^{-1}
\right)
\\
\nonumber
&&+t^{-1}\Big[
V_{ud}\left(Bc_{K1}^{\alpha\beta}-c_{K3}^{\alpha\beta}(t-p_1^2) \right)(k-p_1)_\mu p_\nu
\nonumber
\\
&&~~~~~~~~~
+V_{us}\left( Bc_{\pi1}^{\beta\alpha}-c_{\pi3}^{\beta\alpha}(t-p_2^2)  \right)k_\mu(k-p_1)_\nu\Big]
\nonumber
\\
&&+u^{-1}\Big[
V_{ud}\left(Bc_{K1}^{\beta\alpha}-c_{K3}^{\beta\alpha}(u-p_2^2)\right) p_\mu(k-p_2)_\nu
\nonumber
\\
&&~~~~~~~~~
+V_{us}\left(Bc_{\pi1}^{\alpha\beta}-c_{\pi3}^{\alpha\beta} (u-p_1^2)\right)(k-p_2)_\mu k_\nu \Big],
\\
T_{2\mu\nu\rho}&=&V_{ud}c_{K2}^{\alpha\beta}k_\mu(k-p_1)_\nu p_\rho t^{-1}
-V_{us}c_{\pi2}^{\alpha\beta}p_\mu (k-p_2)_\nu k_\rho u^{-1},
\\
T_{3\mu\nu\rho}&=&V_{us}c_{\pi2}^{\beta\alpha}k_\mu(k-p_1)_\nu p_\rho t^{-1}
- V_{ud}c_{K2}^{\beta\alpha}p_\mu(k-p_2)_\nu k_\rho u^{-1},
\end{eqnarray}
with $s=(p_1+p_2)^2,~t=(k-p_1)^2$, and $u=(k-p_2)^2$. The amplitude has the correct antisymmetry under interchange of the two leptons upon using the relations for bilinear spinor wavefunctions
\begin{eqnarray}
&&\overline{u_\alpha}P_Ru_\beta^C
=-\overline{u_\beta}P_Ru_\alpha^C,~~~
\overline{u_\alpha}\gamma^\mu\gamma^\nu P_Ru_\beta^C
=-\overline{u_\beta}\gamma^\nu\gamma^\mu P_Ru_\alpha^C,
\nonumber
\\
&&\overline{u_\alpha}\gamma^\mu\gamma^\nu\gamma^\rho P_Ru_\beta^C
=\overline{u_\beta}\gamma^\rho\gamma^\nu\gamma^\mu P_Lu_\alpha^C,
\end{eqnarray}
and obvious relations for $T_\textrm{SD}$ and $T_i$ tensors.
The decay width is calculated as
\begin{eqnarray}
\label{decaywidth}
\Gamma=\frac{1}{1+\delta_{\alpha\beta}}\frac{1}{2m_K}\frac{1}{128\pi^3m_K^2}
\int ds \int dt~ |\mathcal{M}|^2,
\label{decaywidth}
\end{eqnarray}
where the first factor removes double counting in phase space integration for identical particles, and the integration domains are
\begin{eqnarray}
&&s\in \left[(m_\alpha+m_\beta)^2,~(m_K-m_\pi)^2\right],
\\
&&t\in \bigg[
(E_2^*+E_3^*)^2-\left(\sqrt{E_2^{*2}-m_\beta^2}+\sqrt{E_3^{*2}-m_\pi^2}
\right)^2,
\nonumber
\\
&&~~~~~~~~
(E_2^*+E_3^*)^2-\left(\sqrt{E_2^{*2}-m_\beta^2}
-\sqrt{E_3^{*2}-m_\pi^2}\right)^2\bigg],
\end{eqnarray}
with $m_{K,\pi,\alpha,\beta}$ being the masses of the $K^-,~\pi^+,~l_\alpha$, and $l_\beta$ respectively and
\begin{eqnarray}
E_2^*=\frac{1}{2\sqrt{s}}(s-m_\alpha^2+m_\beta^2),~~~
E_3^*=\frac{1}{2\sqrt{s}}(m_K^2-s-m_\pi^2).
\end{eqnarray}

Now we make some numerical analysis. The values of the SM parameters are taken from the Particle Data Group~\cite{Tanabashi:2018oca}:
\begin{align}
\nonumber
\Gamma_K^{\text{exp}}&=5.3166\times10^{-14}~{\MeV}, &
G_F&=1.1664\times10^{-5}~{\GeV}^{-2},
\\
V_{ud}&=0.9743, & V_{us}&=0.2253,
\nonumber
\\
m_K&=493.677~{\MeV}, &
m_\pi&=139.570~{\MeV},
\nonumber
\\
m_e&=0.511~{\MeV}, &
m_\mu&=105.658~{\MeV},
\end{align}
together with the $\chpt$ parameters $F_0=87~{\MeV}$~\cite{Colangelo:2003hf} and $B=2.8~{\GeV}$~\cite{Cirigliano:2017djv}. Our master formulae for the branching ratios of the decays $K^-\rightarrow \pi^+l_\alpha^-l_\beta^-$ are
\begin{eqnarray}
\nonumber
{\mathcal{B}(e^-e^-)\over \GeV^6}
&=&\frac{1.7\times 10^{-33}}{\GeV^6}\frac{|m_{ee}|^2}{\eV^2}
+80\left|\calY^{ee}_{K 1}\right|^2
+4.3\left|\calY^{ee}_{\pi 1}\right|^2
\\
\nonumber
&&
+10^{-3}\times\left(48\left|\calX^{ee}_1\right|^2
+45\left|\calY^{ee}_{K 2}\right|^2
+2.4\left|\calY^{ee}_{\pi 2}\right|^2\right)
\\
&&+10^{-8}\times\left(29\left|\calY^{ee}_{K 3}\right|^2
+23\left|\calX^{ee}_2\right|^2
+1.6\left|\calY^{ee}_{\pi 3}\right|^2\right)+\textrm{int.},
\\
\nonumber
{\mathcal{B}(\mu^-\mu^-)\over  \GeV^6}&=&
\frac{4.5\times 10^{-34}}{\GeV^6}\frac{|m_{\mu\mu}|^2}{\eV^2}
+16\left|\calY^{\mu\mu}_{K 1}\right|^2
+2.2\left|\calY^{\mu\mu}_{\pi 1}\right|^2
\\
\nonumber
&&
+10^{-3}\times\left(17\left|\calX^{\mu\mu}_1\right|^2
+19\left|\calY^{\mu\mu}_{K 2}\right|^2
+\left|\calY^{\mu\mu}_{\pi 2}\right|^2\right)
\\
&&+10^{-9}\times\left(67\left|\calX^{\mu\mu}_2\right|^2
+49\left|\calY^{\mu\mu}_{K 3}\right|^2
+6.6\left|\calY^{\mu\mu}_{\pi 3}\right|^2\right)
+\textrm{int.},
\\
\nonumber
{\mathcal{B}(e^-\mu^-)\over  \GeV^6}&=&
\frac{2.1\times 10^{-33}}{\GeV^6}\frac{|m_{e\mu}|^2}{\eV^2}
+26\left|\calY^{\mu e}_{K 1}\right |^2
+17\left|\calY^{e\mu}_{K 1}\right |^2
+2\left|\calY^{e\mu}_{\pi 1}\right |^2
+1.4\left|\calY^{\mu e}_{\pi 1}\right |^2
\\
\nonumber
&&
+10^{-3}\times\left(61\left|\calX^{e\mu}_1\right|^2
+35\left|\calY^{\mu e}_{K 2}\right|^2
+24\left|\calY^{e\mu}_{K 2}\right|^2
+1.9\left|\calY^{e\mu}_{\pi 2}\right|^2
+1.3\left|\calY^{\mu e}_{\pi 2}\right|^2\right)
\\
&&
+10^{-9}\times\left(280\left|\calX^{e\mu}_2\right|^2
+110\left|\calY^{e\mu}_{K 3}\right|^2
+55\left|\calY^{\mu e}_{K 3}\right|^2
+6.7\left|\calY^{\mu e}_{\pi 3}\right|^2
+5.7\left|\calY^{e\mu}_{\pi 3}\right|^2\right)
+\textrm{int.},
\end{eqnarray}
where int. stands for interference terms between any pair of Wilson coefficients, whose complete forms are displayed in appendix~\ref{app2}. We can see a few features from the above results. First of all, since the neutrino mass scale is at most $\calO(\eV)$~\cite{Aker:2019uuj,Loureiro:2018pdz},
the contribution from the neutrino mass matrix is negligible for any measurable branching ratios in a collider-type experiment. Second, if we assume the Wilson coefficients associated with the dim-7 operators in SMEFT are similar in size, all of $\calX$ and $\calY$ parameters will be of a similar order of magnitude. Their relative importance is measured by their prefactors, which have the rough ratios:
\begin{eqnarray}
%\textrm{ratios of prefactors of }
\calY_1:\calX_1:\calY_2:\calX_2:\calY_3\sim
10^{1}:10^{-2}:10^{-3\sim -2}:10^{-7}:10^{-8\sim -7}.
\end{eqnarray}
Generically speaking, a long-distance contribution ($\calY_{P_i,j}$ term) due to a neutrino exchange in figure~\ref{fig1}(b) dominates over its similar short-distance one ($\calX_j$ term) in figure~\ref{fig1}(c), which in turn is similar to a $\calY_{P_i j+1}$ term.

\begin{table}
\centering
\begin{tabular}{| lc| lc| lc| lc|}
\hline
\multicolumn{2}{|c}{$K^-\rightarrow \pi^+e^-e^- $} &  \multicolumn{2}{|c}{ $K^-\rightarrow \pi^+\mu^-\mu^- $}
&\multicolumn{4}{|c|}{ $K^-\rightarrow \pi^+e^-\mu^-$}
\\
\hline
names & bounds &names & bounds &names & bounds &names & bounds
\\
\hline
$\left|\calY^{ee}_{K1}\right |^{-\frac{1}{3}}$ & $84.5$
& $\left|\calY^{\mu\mu}_{K1}\right |^{-\frac{1}{3}}$ & $85.1$
& $\left|\calY^{\mu e}_{K1}\right |^{-\frac{1}{3}}$ & $61.1$
& $\left|\calY^{e\mu}_{K1}\right |^{-\frac{1}{3}}$ & $56.9$
 \\
\hline
$\left|\calY^{ee}_{\pi1}\right |^{-\frac{1}{3}}$ & $51.9$
& $\left|\calY^{\mu\mu}_{\pi1}\right |^{-\frac{1}{3}}$ & $61.2$
& $\left|\calY^{e\mu}_{\pi1}\right |^{-\frac{1}{3}}$ & $39.8$
& $\left|\calY^{\mu e}_{\pi1}\right |^{-\frac{1}{3}}$ & $37.5$
 \\
 \hline
$\left|\calX^{ee}_1\right |^{-\frac{1}{3}}$ & $24.5$
& $\left|\calX^{\mu\mu}_1\right |^{-\frac{1}{3}}$ & $32.3$
&$\left|\calX^{e\mu}_1\right |^{-\frac{1}{3}}$ & $22.3$
& &
\\
\hline
$\left|\calY^{ee}_{K2}\right |^{-\frac{1}{3}}$ & $24.3$
&$\left|\calY^{\mu\mu}_{K2}\right |^{-\frac{1}{3}}$ & $27.7$
&$\left|\calY^{\mu e}_{K2}\right |^{-\frac{1}{3}}$ & $20.3$
&$\left|\calY^{e\mu}_{K2}\right |^{-\frac{1}{3}}$ & $19.1$
 \\
 \hline
$\left|\calY^{ee}_{\pi2}\right |^{-\frac{1}{3}}$ & $14.9$
&$\left|\calY^{\mu\mu}_{\pi2}\right |^{-\frac{1}{3}}$ & $17$
&$\left|\calY^{e\mu}_{\pi2}\right |^{-\frac{1}{3}}$ & $12.5$
&$\left|\calY^{\mu e}_{\pi2}\right |^{-\frac{1}{3}}$ & $11.7$
 \\
 \hline
$\left|\calX^{ee}_2\right |^{-\frac{1}{3}}$ & $3.2$
& $\left|\calX^{\mu\mu}_2\right |^{-\frac{1}{3}}$ & $3.4$
&$\left|\calX^{e\mu}_2\right |^{-\frac{1}{3}}$ & $2.9$
& &
 \\
 \hline
 $\left|\calY^{ee}_{K3}\right |^{-\frac{1}{3}}$ & $3.3$
 & $\left|\calY^{\mu\mu}_{K3}\right |^{-\frac{1}{3}}$ & $3.2$
 & $\left|\calY^{e\mu}_{K3}\right |^{-\frac{1}{3}}$ & $2.6$
& $\left|\calY^{\mu e}_{K3}\right |^{-\frac{1}{3}}$ & $2.2$
 \\
 \hline
  $\left|\calY^{ee}_{\pi3}\right |^{-\frac{1}{3}}$ & $2$
  & $\left|\calY^{\mu\mu}_{\pi3}\right |^{-\frac{1}{3}}$ & $2.3$
  & $\left|\calY^{\mu e}_{\pi3}\right |^{-\frac{1}{3}}$ & $1.5$
  & $\left|\calY^{e\mu}_{\pi3}\right |^{-\frac{1}{3}}$ & $1.5$
  \\
  \hline
\end{tabular}
\caption{Lower bounds (in units of GeV) are shown for inverse cubic roots ($|\calX_i|^{-1/3}$ or $|\calY_i|^{-1/3}$) of combinations of Wilson coefficients for dim-7 operators in SMEFT. Note that $\calX_i^{\alpha\beta}=\calX_i^{\beta\alpha}$.}
\label{tab1}
\end{table}

\begin{figure}[!h]
\centering
\includegraphics[width=12cm]{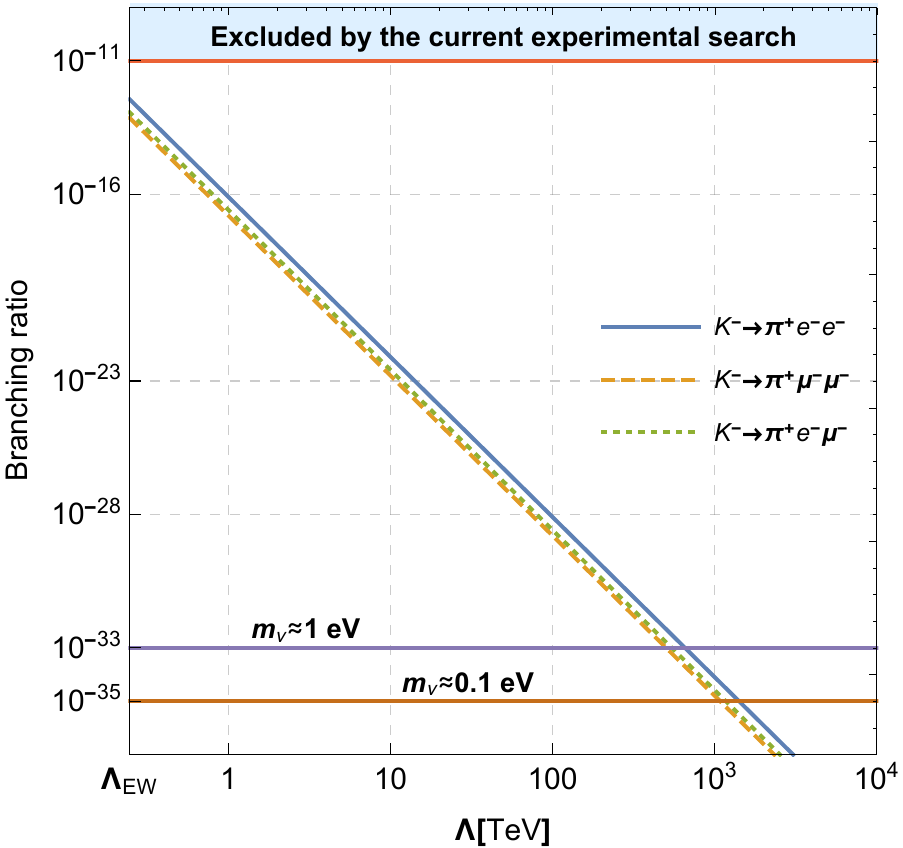}
\caption{Branching ratios for $K^-\to\pi^+l_\alpha^-l_\beta^-$ are shown as a function of the new physics scale $\Lambda$ in the SMEFT framework. Also shown are the current experimental bounds (upper horizontal line) and the neutrino mass contribution alone (lower horizontal lines).}
\label{fig2}
\end{figure}

The current experimental upper bounds on the above decays are
\begin{align}
\mathcal{B}_{\text{exp}}(e^-e^-)&<2.2\times 10^{-10}~\cite{CortinaGil:2019dnd}, &
\mathcal{B}_{\text{exp}}(\mu^-\mu^-)&<4.2\times 10^{-11}~\cite{CortinaGil:2019dnd}, &
\mathcal{B}_{\text{exp}}(e^-\mu^-)&<5\times 10^{-10}~\cite{Appel:2000tc}.
\end{align}
To get some numerical feel on what these bounds would imply and considering the limited number of experimental bounds compared to that of Wilson coefficients, we assume only one of the $\calX_i$ or $\calY_i$ is nonzero. The above upper bounds on branching ratios then translate into the lower bounds on their inverse cubic roots as displayed in table~\ref{tab1}. The bounds are rather weak, especially when compared with those from nuclear $0\nu\beta\beta$ decays~\cite{Cirigliano:2018yza,Liao:2019tep}. This relative weakness originates from much smaller data samples accumulated in kaon experiments than the number of nuclei available in a ton-level experiment of $0\nu\beta\beta$ decays as we estimated roughly in Ref.~\cite{Liao:2019gex}. Thus the weak bounds should not be interpreted as if the SMEFT approach would be valid for a new particle with a mass as low as tens or even a few GeV; on the contrary, if there are such particles, they must be incorporated explicitly into the expanded version of SMEFT and even LEFT. Nevertheless, we stress that however weak the bounds are, they are the first ones worked out in a systematic effective field theory approach that involve the second generation of fermions and are thus complementary to those obtained from nuclear $0\nu\beta\beta$ decays. Conversely, if we assume the dim-7 Wilson coefficients are all of order $\Lambda^{-3}$ where $\Lambda$ is the new physics scale, the branching ratios are dominated by the terms with the largest coefficients, i.e., the long-distance terms of $\calY^{\alpha\beta}_{P_i 1}$. In figure~\ref{fig2} we plot our theoretical predictions as a function of $\Lambda$, together with the current experimental bounds; also shown are the contributions from the neutrino mass matrix alone assuming $m_\nu\approx 0.1\textrm{ or }1~\eV$. For instance, if $\Lambda>1~\TeV$ as the current LHC searches and the null results in nuclear $0\nu\beta\beta$ decays imply, we have for SMEFT: 
\begin{align}
%\textrm{SMEFT with }\Lambda&>1~\TeV: 
\mathcal{B}(e^-e^-)&<8.0\times 10^{-17}, &
\mathcal{B}(\mu^-\mu^-)&<1.6\times 10^{-17}, &
\mathcal{B}(e^-\mu^-)&<2.6\times 10^{-17}.
\end{align}
These branching ratios are several orders of magnitude smaller than the current experimental upper bounds.

%%%%%%%%%%%%%%%%%%%%%%%
\section{Conclusion}
%%%%%%%%%%%%%%%%%%%%%%%
\label{sec5}

We have accomplished a comprehensive analysis on the lepton number violating decays $K^\pm\rightarrow\pi^\mp l^{\pm}_\alpha l^{\pm}_\beta$ in the effective field theory approach. We focused in this work on the long-distance contribution due to an exchange of neutrinos, and incorporated the short-distance contribution obtained in our previous work~\cite{Liao:2019gex}. It turns out that the long-distance contribution overwhelmingly dominates over the short-distance one by about three orders of magnitude in the decay branching ratios. Assuming there are no new particles with a mass of the order of or below the electroweak scale, we related the decay branching ratios to the leading LNV effective interactions in SMEFT due to dim-5 and dim-7 operators. Our results are general in that subject to the above loose assumption they are independent of dynamical details at an even higher new physics scale; instead, different dynamics are hidden in the Wilson coefficients in SMEFT. Our results do not employ any hadronic models, but are completely based on the well-established symmetries and effective field theories from SMEFT through LEFT to $\chi$PT. While the hadronic LECs entering in the short-distance contribution were fixed previously by experimental measurements and lattice calculations, the long-distance contribution involves no parameters other than the pion decay constant and quark condensate making our results very robust. Unfortunately, the current experimental upper bounds on the decay branching ratios are too weak to set a useful bound on the scale of new physics that is responsible for lepton number violation.

\vspace{0.5cm}
\noindent
%%%%%%%%%%%%%%%%%%%%%%%
\section*{Acknowledgement}
%%%%%%%%%%%%%%%%%%%%%%%
We would like to thank Chang-Yuan Yao for discussions on numerical calculations. This work was supported in part by the Grants No.~NSFC-11975130, No.~NSFC-11575089, by The National Key Research and Development Program of China under Grant No. 2017YFA0402200, by the CAS Center for Excellence in Particle Physics (CCEPP). Xiao-Dong Ma is supported by Grant No.~MOST~106-2112-M-002-003-MY3.

%%%%%%%%%%%%%%%%%%%%%%%
%%%%%%%%%%%%%%%%%%%%%%%
\begin{appendices}
\numberwithin{equation}{section}
\setcounter{equation}{0}

%%%%%%%%%%%%%%%%%%%%%%%
\section{Baryon number conserving dim-7 operators in SMEFT}
%%%%%%%%%%%%%%%%%%%%%%%
\label{app1}
\begin{table}[!ht]
\centering
\begin{tabular}{|c|c|c|c|}
 \multicolumn{2}{c}{$\psi^2H^4+\mbox{h.c.}$} &  \multicolumn{2}{c}{ $\psi^2H^3D+\mbox{h.c.}$}
\\
\hline
$\mathcal{O}_{LH}$ & $\epsilon_{ij}\epsilon_{mn}(\overline{L^{C,i}}L^m)H^jH^n(H^\dagger H)$ & $\mathcal{O}_{LeHD}$ & $\epsilon_{ij}\epsilon_{mn}(\overline{L^{C,i}}\gamma_\mu e)H^j(H^miD^\mu H^n)$
\\
\hline
\multicolumn{2}{c}{$\psi^2H^2D^2+\mbox{h.c.}$}&  \multicolumn{2}{c}{$\psi^2H^2X+\mbox{h.c.}$}
\\
\hline
$\mathcal{O}_{LDH1}(*)$ & $\epsilon_{ij}\epsilon_{mn}(\overline{L^{C,i}}\overleftrightarrow{D}_\mu L^j)(H^mD^\mu H^n)$ &
$\mathcal{O}_{LHB}$ & $ g_1\epsilon_{ij}\epsilon_{mn}(\overline{L^{C,i}}\sigma_{\mu\nu}L^m)H^jH^nB^{\mu\nu}$
\\
$\mathcal{O}_{LDH2}(*)$ & $\epsilon_{im}\epsilon_{jn}(\overline{L^{C,i}}L^j)(D_\mu H^m D^\mu  H^n)$ &
$\mathcal{O}_{LHW}$ & $g_2\epsilon_{ij}(\epsilon \tau^I)_{mn}(\overline{L^{C,i}}\sigma_{\mu\nu}L^m)H^jH^nW^{I\mu\nu}$
\\
\hline
\multicolumn{2}{c}{$\psi^4D+\mbox{h.c.}$}  &   \multicolumn{2}{c}{$\psi^4H+\mbox{h.c.}$}
\\
\hline
$\mathcal{O}_{\overline{d}uLDL}(*)$ & $\epsilon_{ij}(\overline{d}\gamma_\mu u)(\overline{L^{C,i}}i\overleftrightarrow{D}^\mu L^j)$ &
$\mathcal{O}_{\overline{e}LLLH}$ & $\epsilon_{ij}\epsilon_{mn}(\overline{e}L^i)(\overline{L^{C,j}}L^m)H^n$
\\
&  &
$\mathcal{O}_{\overline{d}QLLH1}(*)$ & $\epsilon_{ij}\epsilon_{mn}(\overline{d}Q^i)(\overline{L^{C,j}}L^m)H^n$
\\
&  &
$\mathcal{O}_{\overline{d}QLLH2}(*)$ & $\epsilon_{ij}\epsilon_{mn}(\overline{d}\sigma_{\mu\nu}Q^i)(\overline{L^{C,j}}\sigma^{\mu\nu} L^m)H^n$
\\
&  &
$\mathcal{O}_{\overline{d}uLeH}(*)$ & $\epsilon_{ij}(\overline{d}\gamma_\mu u)(\overline{L^{C,i}}\gamma^\mu e)H^j$
\\
&  &
$\mathcal{O}_{\overline{Q}uLLH}$ & $\epsilon_{ij}(\overline{Q}u)(\overline{L^{C}}L^i)H^j$
\\
\hline
\end{tabular}
\caption{Basis of dim-7 lepton number violating but baryon number conserving operators in SMEFT. $L,~Q$ are the SM left-handed lepton and quark doublet fields, $u,~d,~e$ are the right-handed up-type quark, down-type quark and charged lepton singlet fields, and $H$ denotes the Higgs doublet, respectively. $D^\mu H^n$ is understood as $(D^\mu H)^n$.}
\label{tab2}
\end{table}

The dim-5 operator in SMEFT is well-known~\cite{Weinberg:1979sa}:
\begin{eqnarray}
\calO_5=\epsilon_{ij}\epsilon_{mn}(\overline{L^{C,i}}L^m)H^jH^n.
\end{eqnarray}
The dim-7 operators were first systematically studied in Ref.~\cite{Lehman:2014jma}, and corrected by Ref.~\cite{Liao:2016hru}. In this appendix we improve further over the basis of operators in Ref.~\cite{Liao:2016hru} so that flavor symmetries are apparently realized as advocated in Ref.~\cite{Liao:2019tep}. This only concerns the subset of operators that violate lepton number but conserve baryon number. In table~\ref{tab2} the newly chosen basis operators are indicated by $(*)$, which replace the following old basis operators~\cite{Liao:2016hru}:
\begin{eqnarray}
\mathcal{O}_{LHD1}^{pr}&=&
\epsilon_{ij}\epsilon_{mn}(\overline{L^{C,i}_p}D_\mu L^j_r)(H^mD^\mu H^n), 
\\
\mathcal{O}_{LHD2}^{pr}&=&
\epsilon_{im}\epsilon_{jn}(\overline{L^{C,i}_p}D_\mu L^j_r)(H^mD^\mu H^n),
\\
\mathcal{O}_{\bar{d}uLLD}^{prst}&=&
\epsilon_{ij}(\overline{d_p}\gamma_\mu u_r)(\overline{L^{C,i}_s}iD^\mu  L^j_t), 
\\
\mathcal{O}_{\bar{d}LQLH1}^{prst}&=&\epsilon_{ij}\epsilon_{mn}
(\overline{d_p}L_r^i)(\overline{Q^{C,j}_s}L^m_t)H^n,
\\
\mathcal{O}_{\bar{d}LueH}^{prst}&=&
\epsilon_{ij}(\overline{d_p}L_r^i)(\overline{u^{C}_s}e_t)H^j, 
\\
\mathcal{O}_{\bar{d}LQLH2}^{prst}&=&
\epsilon_{im}\epsilon_{jn}
(\overline{d_p}L_r^i)(\overline{Q^{C,j}_s}L^m_t)H^n.
\end{eqnarray}
The relations between the new (left) and old (right) operators are as follows,
\begin{eqnarray}
\mathcal{O}_{LDH1}^{pr}&=&
\mathcal{O}_{LHD1}^{pr}+\mathcal{O}_{LHD1}^{rp}, 
\\
\mathcal{O}_{LDH2}^{pr}&=&
-\left( \mathcal{O}_{LHD2}^{pr}+\mathcal{O}_{LHD2}^{rp}\right)
+\frac{1}{2}\left(\mathcal{O}_{LHD1}^{pr}+\mathcal{O}_{LHD1}^{rp}\right)
+\textrm{EoM},
\\
\mathcal{O}_{\bar{d}uLDL}^{prst}&=&
\mathcal{O}_{\bar{d}uLLD}^{prst}+\mathcal{O}_{\bar{d}uLLD}^{prts}, 
\\
\mathcal{O}_{\bar{d}QLLH1}^{prst}&=&
\mathcal{O}_{\bar{d}LQLH1}^{psrt}+\mathcal{O}_{\bar{d}LQLH1}^{ptrs}
-\mathcal{O}_{\bar{d}LQLH2}^{ptrs},
\\
\mathcal{O}_{\bar{d}uLeH}^{prst}&=&
2\mathcal{O}_{\bar{d}LueH}^{psrt},  
\\
\mathcal{O}_{\bar{d}QLLH2}^{prst}&=&
-4\left(\mathcal{O}_{\bar{d}LQLH1}^{psrt}
-\mathcal{O}_{\bar{d}LQLH1}^{ptrs}+ \mathcal{O}_{\bar{d}LQLH2}^{ptrs}\right),
\end{eqnarray}
where EoM refers to equations of motion terms.

%%%%%%%%%%%%%%%%%%%%%%%
\section{Complete results for branching ratios in SMEFT }
%%%%%%%%%%%%%%%%%%%%%%%
\label{app2}

In this Appendix we show the complete results for the branching ratios in terms of the Wilson coefficients in SMEFT:
\begin{eqnarray}%%%%%
&&{\mathcal{B}(e^-e^-)\over \GeV^6}
\\
&=&
2.3\times10^4v^{-8}\left|m_{ee}\right |^2
+2.7 \times10^3v^{-4} \Re\left(m_{ee}\calY^{ee*}_{K1}\right)
+6.3\times10^2v^{-4}\Re\left(m_{ee}\calY^{ee*}_{\pi1}\right)
\nonumber
\\
&&
+8\times10  \left|\calY^{ee}_{K1}\right |^2
-6.6\times10v^{-4}\Re\left(m_{ee}\calX^{ee*}_1\right)
+3.7\times10\Re\left(\calY^{ee}_{K1}\calY^{ee*}_{\pi1}\right)
+4.3 \left|\calY^{ee}_{\pi1}\right |^2
\nonumber
\\
\nonumber
&&
-3.9\Re\left(\calY^{ee}_{K1}\calX^{ee*}_1\right)
-9\times10^{-1}\Re\left(\calY^{ee}_{\pi1}\calX^{ee*}_1\right)
+4.2\times10^{-1}v^{-4}\Re\left(m_{ee}\calY^{ee*}_{K2}\right)
\\\nonumber
&&
-1.6\times10^{-1}v^{-4}\Re\left(m_{ee}\calY^{ee*}_{K3}\right)
-1.4\times10^{-1}v^{-4}\Re\left(m_{ee}\calX^{ee*}_2\right)
-6.3\times10^{-2}v^{-4}\Re\left(m_{ee}\calY^{ee*}_{\pi2}\right)
\\\nonumber
&&
+4.8\times10^{-2}\left|\calX^{ee}_1\right|^2
+4.5\times10^{-2}\left|\calY^{ee}_{K2}\right |^2
-3.8\times10^{-2}v^{-4}\Re\left(m_{ee}\calY^{ee*}_{\pi3}\right)
\\\nonumber
&&
-2.1\times10^{-2}\Re\left(\calY^{ee}_{K2}\calY^{ee*}_{\pi2}\right)
+2.0\times10^{-2}\Re\left(\calY^{ee}_{K1}\calY^{ee*}_{K2}\right)
-9.6\times 10^{-3}\Re\left(\calY^{ee}_{K1}\calY^{ee*}_{K3}\right)
\\\nonumber
&&
-8.6\times10^{-3}\Re\left(\calY^{ee}_{K1}\calX^{ee*}_2\right)
+6.8\times10^{-3}\Re\left(\calY^{ee}_{\pi1}\calY^{ee*}_{K2}\right)
-2.7\times 10^{-3}\Re\left(\calY^{ee}_{K1}\calY^{ee*}_{\pi2}\right)
\\\nonumber
&&
+2.4\times10^{-3}\left|\calY^{ee}_{\pi2} \right|^2
-2.2\times10^{-3}\Re\left(\calY^{ee}_{\pi1}\calY^{ee*}_{K3}\right)
-2.2\times10^{-3}\Re\left(\calY^{ee}_{K1}\calY^{ee*}_{\pi3}\right)
\\\nonumber
&&
-2.0\times10^{-3}\Re\left(\calY^{ee}_{\pi1}\calX^{ee*}_2\right)
-1.1\times10^{-3}\Re\left(\calY^{ee}_{\pi1}\calY^{ee*}_{\pi2}\right)
-6\times10^{-4}\Re\left(\calY^{ee}_{K2}\calX^{ee*}_1\right)
\\\nonumber
&&
-5.1\times10^{-4}\Re\left(\calY^{ee}_{\pi1}\calY^{ee*}_{\pi3}\right)
+2.3\times10^{-4}\Re\left(\calX^{ee}_1\calY^{ee*}_{K3}\right)
+2.1\times10^{-4}\Re\left(\calX^{ee}_1\calX^{ee*}_2\right)
\\\nonumber
&&
+9\times10^{-5}\Re\left(\calY^{ee}_{\pi2}\calX^{ee*}_1\right)
+5.4\times10^{-5}\Re\left(\calX^{ee}_1\calY^{ee*}_{\pi3}\right)
-1.4\times10^{-6}\Re\left(\calY^{ee}_{K2}\calX^{ee*}_2\right)
\\\nonumber
&&
-1.4\times10^{-6}\Re\left(\calY^{ee}_{K2}\calY^{ee*}_{K3}\right)
+5.2\times10^{-7}\Re\left(\calY^{ee}_{K3}\calX^{ee*}_2\right)
-4.4\times10^{-7}\Re\left(\calY^{ee}_{K2}\calY^{ee*}_{\pi3}\right)
\\\nonumber
&&
+2.9\times10^{-7}\left|\calY^{ee}_{K3} \right|^2
+2.3\times10^{-7}\left|\calX^{ee}_2\right|^2
+2.3\times10^{-7}\Re\left(\calX^{ee}_2\calY^{ee*}_{\pi2}\right)
\\\nonumber
&&
+1.9\times10^{-7}\Re\left(\calY^{ee}_{\pi2}\calY^{ee*}_{K3}\right)
+1.4\times10^{-7}\Re\left(\calY^{ee}_{K3}\calY^{ee*}_{\pi3}\right)
+1.2\times10^{-7}\Re\left(\calX^{ee}_2\calY^{ee*}_{\pi3}\right)
\\\nonumber
&&
+7.3\times10^{-8}\Re\left(\calY^{ee}_{\pi2}\calY^{ee*}_{\pi3}\right)
+1.6\times10^{-8}\left|\calY^{ee}_{\pi3}\right|^2
\\%%%%%
&&{\mathcal{B}(\mu^-\mu^-)\over \GeV^6}
\\
&=&
6.1\times10^3v^{-8}\left|m_{\mu\mu}\right |^2
+6.1\times10^2v^{-4} \Re\left(m_{\mu\mu}\calY^{\mu\mu*}_{K1}\right)
+2.3\times10^{2}v^{-4}\Re\left(m_{\mu\mu}\calY^{\mu\mu*}_{\pi1}\right)
\nonumber
\\
\nonumber
&&
-2\times10v^{-4}\Re\left(m_{\mu\mu}\calX^{\mu\mu*}_1\right)
+1.9\times10v^{-4}\Re\left(m_{\mu\mu}\calY^{\mu\mu*}_{K2}\right)
+1.6\times10\left|\calY^{\mu\mu}_{K1}\right|^2
\\\nonumber
&&
+1.1\times10\Re\left(\calY^{\mu\mu}_{K1}\calY^{\mu\mu*}_{\pi1}\right)
-3.2v^{-4}\Re\left(m_{\mu\mu}\calY^{\mu\mu*}_{\pi2}\right)
+2.2\left|\calY^{\mu\mu}_{\pi1}\right|^2
-\Re\left(\calY^{\mu\mu}_{K1}\calX^{\mu\mu*}_1\right)
\\\nonumber
&&
+8.4\times10^{-1}\Re\left(\calY^{\mu\mu}_{K1}\calY^{\mu\mu*}_{K2}\right)
+3.8\times10^{-1}\Re\left(\calY^{\mu\mu}_{\pi1}\calY^{\mu\mu*}_{K2}\right)
-3.8\times10^{-1}\Re\left(\calY^{\mu\mu}_{\pi1}\calX^{\mu\mu*}_1\right)
\\\nonumber
&&
-1.2\times10^{-1}\Re\left(\calY^{\mu\mu}_{K1}\calY^{\mu\mu*}_{\pi2}\right)
-7.1\times10^{-2}\Re\left(\calY^{\mu\mu}_{\pi1}\calY^{\mu\mu*}_{\pi2}\right)
-4\times10^{-2}v^{-4}\Re\left(m_{\mu\mu}\calX^{\mu\mu*}_2\right)
\\\nonumber
&&
-3.4\times10^{-2}v^{-4}\Re\left(m_{\mu\mu}\calY^{\mu\mu*}_{K3}\right)
-3.1\times10^{-2}\Re\left(\calX^{\mu\mu}_1\calY^{\mu\mu*}_{K2}\right)
+1.9\times10^{-2}\left|\calY^{\mu\mu}_{K2}\right|^2
\\\nonumber
&&
+1.7\times10^{-2}\left|\calX^{\mu\mu}_1\right|^2
-1.2\times10^{-2}v^{-4}\Re\left(m_{\mu\mu}\calY^{\mu\mu*}_{\pi3}\right)
-7.9\times10^{-3}\Re\left(\calY^{\mu\mu*}_{\pi2}\calY^{\mu\mu}_{K2}\right)
\\\nonumber
&&
+5.1\times10^{-3}\Re\left(\calY^{\mu\mu}_{\pi2}\calX^{\mu\mu*}_1\right)
-2.0\times10^{-3}\Re\left(\calY^{\mu\mu}_{K1}\calX^{\mu\mu*}_2\right)
-1.7\times10^{-3}\Re\left(\calY^{\mu\mu}_{K1}\calY^{\mu\mu*}_{K3}\right)
\\\nonumber
&&
+1\times10^{-3}\left|\calY^{\mu\mu}_{\pi2}\right|^2
-7.5\times10^{-4}\Re\left(\calY^{\mu\mu}_{\pi1}\calX^{\mu\mu*}_2\right)
-6.1\times10^{-4}\Re\left(\calY^{\mu\mu}_{\pi1}\calY^{\mu\mu*}_{K3}\right)
\\\nonumber
&&
-6\times10^{-4}\Re\left(\calY^{\mu\mu}_{K1}\calY^{\mu\mu*}_{\pi3}\right)
-2.4\times10^{-4}\Re\left(\calY^{\mu\mu}_{\pi1}\calY^{\mu\mu*}_{\pi3}\right)
+6.7\times10^{-5}\Re\left(\calX^{\mu\mu}_1\calX^{\mu\mu*}_2\right)
\\\nonumber
&&
-6.2\times10^{-5}\Re\left(\calY^{\mu\mu}_{K2}\calX^{\mu\mu*}_2\right)
+5.6\times10^{-5}\Re\left(\calX^{\mu\mu}_1\calY^{\mu\mu*}_{K3}\right)
-4.7\times10^{-5}\Re\left(\calY^{\mu\mu}_{K2}\calY^{\mu\mu*}_{K3}\right)
\\\nonumber
&&
-2.1\times10^{-5}\Re\left(\calY^{\mu\mu}_{K2}\calY^{\mu\mu*}_{\pi3}\right)
+2\times10^{-5}\Re\left(\calX^{\mu\mu}_1\calY^{\mu\mu*}_{\pi3}\right)
+1.1\times10^{-5}\Re\left(\calY^{\mu\mu}_{\pi2}\calX^{\mu\mu*}_2\right)
\\\nonumber
&&
+7.2\times10^{-6}\Re\left(\calY^{\mu\mu}_{\pi2}\calY^{\mu\mu*}_{K3}\right)
+4.0\times10^{-6}\Re\left(\calY^{\mu\mu}_{\pi2}\calY^{\mu\mu*}_{\pi3}\right)
+1.1\times10^{-7}\Re\left(\calX^{\mu\mu}_2\calY^{\mu\mu*}_{K3}\right)
\\\nonumber
&&
+6.7\times10^{-8}\left|\calX^{\mu\mu}_2\right|^2
+4.9\times10^{-8}\left|\calY^{\mu\mu}_{K3}\right|^2
+4.1\times10^{-8}\Re\left(\calX^{\mu\mu}_2\calY^{\mu\mu*}_{\pi3}\right)
\\\nonumber
&&
+3.4\times10^{-8}\Re\left(\calY^{\mu\mu}_{K3}\calY^{\mu\mu*}_{\pi3}\right)
+6.6\times10^{-9}\left|\calY^{\mu\mu}_{\pi3}\right|^2,
\\%%%%%
&&{\mathcal{B}(e^-\mu^-)\over \GeV^6}
\\
&=&
2.8\times10^4v^{-8}\left|m_{e\mu}\right |^2
+1.7\times10^3v^{-4} \Re\left(m_{e\mu}\calY^{\mu e*}_{K1}\right)
+1.3\times10^3v^{-4}\Re\left(m_{e\mu}\calY^{e\mu*}_{K1}\right)
\nonumber
\\
\nonumber
&&
+4.7\times10^2v^{-4} \Re\left(m_{e\mu}\calY^{e\mu*}_{\pi1}\right)
+3.9\times10^2v^{-4} \Re\left(m_{e\mu}\calY^{\mu e*}_{\pi1}\right)
-8.2\times10v^{-4} \Re\left(m_{e\mu}\calX^{e\mu*}_1\right)
\\\nonumber
&&
+4.5\times10v^{-4} \Re\left(m_{e\mu}\calY^{\mu e*}_{K2}\right)
+4\times10 \Re\left(\calY^{\mu e}_{K1}\calY^{e\mu*}_{K1}\right)
+2.6\times10\left|\calY^{\mu e}_{K1}\right |^2
+1.7\times10\left|\calY^{e\mu}_{K1}\right |^2
\\\nonumber
&&
+1.4\times10 \Re\left(\calY^{\mu e}_{K1}\calY^{e\mu*}_{\pi1}\right)
+1.2\times10 \Re\left(\calY^{\mu e}_{K1}\calY^{\mu e*}_{\pi1}\right)
+9.9\Re\left(\calY^{e\mu}_{K1}\calY^{e\mu*}_{\pi1}\right)
\\\nonumber
&&
+9.3 \Re\left(\calY^{e\mu}_{K1}\calY^{\mu e*}_{\pi1}\right)
-5.9v^{-4}\Re\left(m_{e\mu}\calY^{\mu e*}_{\pi2}\right)
+3.2 \Re\left(\calY^{e\mu}_{\pi1}\calY^{\mu e*}_{\pi1}\right)
\\\nonumber
&&
-2.5 \Re\left(\calY^{\mu e}_{K1}\calX^{e\mu*}_1\right)
+2\left|\calY^{e\mu}_{\pi1}\right |^2
-1.9  \Re\left(\calY^{e\mu}_{K1}\calX^{e\mu*}_1\right)
+1.4\left|\calY^{\mu e}_{\pi1}\right |^2
+1.3 \Re\left(\calY^{\mu e}_{K1}\calY^{\mu e*}_{K2}\right)
\\\nonumber
&&
+6.9\times10^{-1}  \Re\left(\calY^{e\mu}_{K1}\calY^{\mu e*}_{K2}\right)
-6.7\times10^{-1}  \Re\left(\calX^{e\mu}_1\calY^{e\mu*}_{\pi1}\right)
-5.8\times10^{-1}  \Re\left(\calX^{e\mu}_1\calY^{\mu e*}_{\pi1}\right)
\\\nonumber
&&
+4.6\times10^{-1}  \Re\left(\calY^{e\mu}_{\pi1}\calY^{\mu e*}_{K2}\right)
+3\times10^{-1} \Re\left(\calY^{\mu e}_{\pi1}\calY^{\mu e*}_{K2}\right)
+2.4\times10^{-1}v^{-4}\Re\left(m_{e\mu}\calY^{e\mu*}_{K2}\right)
\\\nonumber
&&
-1.7\times10^{-1}v^{-4}  \Re\left(m_{e\mu}\calX^{e\mu*}_2\right)
-1.6\times10^{-1} \Re\left(\calY^{\mu e}_{K1}\calY^{\mu e*}_{\pi2}\right)
-9.7\times10^{-2}v^{-4}\Re\left(m_{e\mu}\calY^{e\mu*}_{K3}\right)
\\\nonumber
&&
-7\times10^{-2}v^{-4}\Re\left(m_{e\mu}\calY^{\mu e*}_{K3}\right)
-6.9\times10^{-2} \Re\left(\calY^{e\mu}_{\pi1}\calY^{\mu e*}_{\pi2}\right)
-6.3\times10^{-2}  \Re\left(\calX^{e\mu}_1\calY^{\mu e*}_{K2}\right)
\\\nonumber
&&
+6.1\times10^{-2}\left| \calX^{e\mu}_1\right|^2
-5.5\times10^{-2} \Re\left(\calY^{e\mu}_{K1}\calY^{\mu e*}_{\pi2}\right)
-4.6\times10^{-2}v^{-4}\Re\left(m_{e\mu}\calY^{e\mu*}_{\pi2}\right)
\\\nonumber
&&
-3.7\times10^{-2} \Re\left(\calY^{\mu e}_{\pi1}\calY^{\mu e*}_{\pi2}\right)
+3.5\times10^{-2}\left| \calY^{\mu e}_{K2}\right|^2
-2.6\times10^{-2}v^{-4}\Re\left(m_{e\mu}\calY^{\mu e*}_{\pi3}\right)
\\\nonumber
&&
+2.4\times10^{-2}\left| \calY^{e\mu}_{K2}\right|^2
-2.3\times10^{-2}v^{-4}\Re\left(m_{e\mu}\calY^{e\mu*}_{\pi3}\right)
-1.3\times10^{-2}  \Re\left(\calY^{e\mu}_{K2}\calY^{e\mu*}_{\pi2}\right)
\\\nonumber
&&
-1.3\times10^{-2}  \Re\left(\calY^{\mu e}_{K2}\calY^{\mu e*}_{\pi2}\right)
+7.8\times10^{-3}  \Re\left(\calX^{e\mu}_1\calY^{\mu e*}_{\pi2}\right)
+6.8\times10^{-3}  \Re\left(\calY^{e\mu}_{K1}\calY^{e\mu*}_{K2}\right)
\\\nonumber
&&
-5.3\times10^{-3}  \Re\left(\calX^{e\mu}_2\calY^{\mu e*}_{K1}\right)
+5\times10^{-3}  \Re\left(\calY^{\mu e}_{K1}\calY^{e\mu*}_{K2}\right)
-4.1\times10^{-3}  \Re\left(\calX^{e\mu}_2\calY^{e\mu*}_{K1}\right)
\\\nonumber
&&
-3\times10^{-3}  \Re\left(\calY^{\mu e}_{K1}\calY^{e\mu*}_{K3}\right)
-2.6\times10^{-3}  \Re\left(\calY^{e\mu}_{K1}\calY^{e\mu*}_{K3}\right)
+2.3\times10^{-3} \Re\left(\calY^{\mu e}_{\pi1}\calY^{e\mu*}_{K2}\right)
\\\nonumber
&&
-2.2\times10^{-3}  \Re\left(\calY^{\mu e}_{K1}\calY^{\mu e*}_{K3}\right)
+2\times10^{-3}  \Re\left(\calY^{e\mu}_{\pi1}\calY^{e\mu*}_{K2}\right)
+1.9\times10^{-3}\left| \calY^{e\mu}_{\pi2}\right|^2
\\\nonumber
&&
-1.6\times10^{-3} \Re\left(\calY^{e\mu}_{K1}\calY^{\mu e*}_{K3}\right)
-1.4\times10^{-3} \Re\left(\calX^{e\mu}_2\calY^{e\mu*}_{\pi1}\right)
+1.3\times10^{-3}\left| \calY^{\mu e}_{\pi2}\right|^2
\\\nonumber
&&
-1.3\times 10^{-3}\Re\left(\calY^{e\mu}_{K1}\calY^{e\mu*}_{\pi2}\right)
-1.2\times10^{-3}  \Re\left(\calY^{\mu e}_{\pi1}\calX^{e\mu*}_2\right)
-7.9\times 10^{-4}\Re\left(\calY^{\mu e}_{K1}\calY^{\mu e*}_{\pi3}\right)
\\\nonumber
&&
-7.4\times 10^{-4}\Re\left(\calY^{e\mu}_{\pi1}\calY^{e\mu*}_{K3}\right)
-7.3\times 10^{-4}\Re\left(\calY^{\mu e}_{K1}\calY^{e\mu*}_{\pi2}\right)
-6.9\times 10^{-4}\Re\left(\calY^{\mu e}_{\pi1}\calY^{e\mu*}_{K3}\right)
\\\nonumber
&&
-6.9\times 10^{-4}\Re\left(\calY^{\mu e}_{K1}\calY^{e\mu*}_{\pi3}\right)
-6.3\times 10^{-4}\Re\left(\calY^{e\mu}_{K1}\calY^{\mu e*}_{\pi3}\right)
-5.9\times 10^{-4}\Re\left(\calY^{e\mu}_{\pi1}\calY^{\mu e*}_{K3}\right)
\\\nonumber
&&
-5.1\times 10^{-4}\Re\left(\calY^{\mu e}_{\pi1}\calY^{\mu e*}_{K3}\right)
-5\times 10^{-4}\Re\left(\calY^{\mu e}_{\pi1}\calY^{e\mu*}_{\pi2}\right)
-4.6\times 10^{-4}\Re\left(\calY^{e\mu}_{K1}\calY^{e\mu*}_{\pi3}\right)
\\\nonumber
&&
-4.5\times 10^{-4}\Re\left(\calY^{e\mu}_{\pi1}\calY^{e\mu*}_{\pi2}\right)
-3.8\times 10^{-4}\Re\left(\calX^{e\mu}_1\calY^{e\mu*}_{K2}\right)
+2.6\times 10^{-4}\Re\left(\calX^{e\mu}_1\calX^{e\mu*}_2\right)
\\\nonumber
&&
-2.1\times 10^{-4}\Re\left(\calY^{e\mu}_{\pi1}\calY^{\mu e*}_{\pi3}\right)
-2\times 10^{-4}\Re\left(\calY^{e\mu}_{\pi1}\calY^{e\mu*}_{\pi3}\right)
-1.8\times 10^{-4}\Re\left(\calY^{\mu e}_{\pi1}\calY^{\mu e*}_{\pi3}\right)
\\\nonumber
&&
-1.6\times 10^{-4}\Re\left(\calY^{\mu e}_{\pi1}\calY^{e\mu*}_{\pi3}\right)
+1.5\times 10^{-4}\Re\left(\calX^{e\mu}_1\calY^{e\mu*}_{K3}\right)
-1.4\times 10^{-4}\Re\left(\calY^{\mu e}_{K2}\calX^{e\mu*}_2\right)
\\\nonumber
&&
+1.1\times 10^{-4}\Re\left(\calX^{e\mu}_1\calY^{\mu e*}_{K3}\right)
+9.5\times 10^{-5}\Re\left(\calY^{e\mu}_{K2}\calY^{\mu e*}_{K2}\right)
+7.3\times 10^{-5}\Re\left(\calX^{e\mu}_1\calY^{e\mu*}_{\pi2}\right)
\\\nonumber
&&
-6.2\times 10^{-5}\Re\left(\calY^{\mu e}_{K2}\calY^{\mu e*}_{K3}\right)
-5\times 10^{-5}\Re\left(\calY^{\mu e}_{K2}\calY^{e\mu*}_{K3}\right)
+3.8\times 10^{-5}\Re\left(\calX^{e\mu}_1\calY^{\mu e*}_{\pi3}\right)
\\\nonumber
&&
+3.4\times 10^{-5}\Re\left(\calX^{e\mu}_1\calY^{e\mu*}_{\pi3}\right)
-2.4\times 10^{-5}\Re\left(\calY^{\mu e}_{K2}\calY^{e\mu*}_{\pi3}\right)
-2.1\times 10^{-5}\Re\left(\calY^{\mu e}_{K2}\calY^{e\mu*}_{\pi2}\right)
\\\nonumber
&&
-1.9\times 10^{-5}\Re\left(\calY^{\mu e}_{K2}\calY^{\mu e*}_{\pi3}\right)
+1.8\times 10^{-5}\Re\left(\calY^{\mu e}_{\pi2}\calX^{e\mu*}_2\right)
-1.2\times 10^{-5}\Re\left(\calY^{e\mu}_{K2}\calY^{\mu e*}_{\pi2}\right)
\\\nonumber
&&
+8.5\times 10^{-6}\Re\left(\calY^{\mu e}_{\pi2}\calY^{\mu e*}_{K3}\right)
+5.1\times 10^{-6}\Re\left(\calY^{e\mu}_{\pi2}\calY^{\mu e*}_{\pi2}\right)
+3.8\times 10^{-6}\Re\left(\calY^{\mu e}_{\pi2}\calY^{e\mu*}_{\pi3}\right)
\\\nonumber
&&
+3.6\times 10^{-6}\Re\left(\calY^{\mu e}_{\pi2}\calY^{e\mu*}_{K3}\right)
+2.4\times 10^{-6}\Re\left(\calY^{\mu e}_{\pi2}\calY^{\mu e*}_{\pi3}\right)
-8.7\times 10^{-7}\Re\left(\calX^{e\mu}_2\calY^{e\mu*}_{K2}\right)
\\\nonumber
&&
-6.1\times 10^{-7}\Re\left(\calY^{e\mu}_{K2}\calY^{e\mu*}_{K3}\right)
+3.1\times 10^{-7}\Re\left(\calX^{e\mu}_2\calY^{e\mu*}_{K3}\right)
+2.8\times 10^{-7}\left|\calX^{e\mu}_2\right|^2
\\\nonumber
&&
+2.3\times 10^{-7}\Re\left(\calX^{e\mu}_2\calY^{\mu e*}_{K3}\right)
-1.8\times 10^{-7}\Re\left(\calY^{e\mu}_{K2}\calY^{\mu e*}_{\pi3}\right)
-1.7\times 10^{-7}\Re\left(\calY^{e\mu}_{K2}\calY^{\mu e*}_{K3}\right)
\\\nonumber
&&
+1.7\times 10^{-7}\Re\left(\calX^{e\mu}_2\calY^{e\mu*}_{\pi2}\right)
+1.2\times 10^{-7}\Re\left(\calY^{e\mu}_{\pi2}\calY^{e\mu*}_{K3}\right)
+1.1\times 10^{-7}\Re\left(\calY^{e\mu}_{K3}\calY^{\mu e*}_{K3}\right)
\\\nonumber
&&
+1.1\times 10^{-7}\left|\calY^{e\mu}_{K3}\right|^2
-9.1\times 10^{-8}\Re\left(\calY^{e\mu}_{K2}\calY^{e\mu*}_{\pi3}\right)
+8.3\times 10^{-8}\Re\left(\calX^{e\mu}_2\calY^{\mu e*}_{\pi3}\right)
\\\nonumber
&&
+7.2\times 10^{-8}\Re\left(\calX^{e\mu}_2\calY^{e\mu*}_{\pi3}\right)
+5.5\times 10^{-8}\left|\calY^{\mu e}_{K3}\right|^2
+5.1\times 10^{-8}\Re\left(\calY^{e\mu}_{K3}\calY^{\mu e*}_{\pi3}\right)
\\\nonumber
&&
+3.9\times 10^{-8}\Re\left(\calY^{e\mu}_{\pi2}\calY^{\mu e*}_{\pi3}\right)
+3.4\times 10^{-8}\Re\left(\calY^{\mu e}_{K3}\calY^{e\mu*}_{\pi3}\right)
+3.2\times 10^{-8}\Re\left(\calY^{e\mu}_{K3}\calY^{e\mu*}_{\pi3}\right)
\\\nonumber
&&
+3\times 10^{-8}\Re\left(\calY^{\mu e}_{K3} \calY^{\mu e*}_{\pi3}\right)
+2.2\times 10^{-8}\Re\left(\calY^{e\mu}_{\pi2}\calY^{\mu e*}_{K3}\right)
+1.9\times 10^{-8}\Re\left(\calY^{e\mu}_{\pi2}\calY^{e\mu*}_{\pi3}\right)
\\\nonumber
&&
+9.6 \times 10^{-9}\Re\left(\calY^{e\mu}_{\pi3} \calY^{\mu e*}_{\pi3}\right)
+6.7 \times 10^{-9}\left| \calY^{\mu e}_{\pi3}\right|^2
+5.7 \times 10^{-9}\left|\calY^{e\mu}_{\pi3}\right|^2.
\end{eqnarray}
Some interference terms have a smaller coefficient than their separate terms because of phase space integration.

\end{appendices}

%%%%%

\end{document}